\documentclass[aps,prc,showpacs]{revtex4}
\usepackage{graphicx}
\usepackage[dvips]{color}
\usepackage{colordvi}

\begin{document}
\bibliographystyle{num}
\title{Correlations and effective interactions in nuclear matter}
\author{     P. Bo\.{z}ek\footnote{Electronic address~:
piotr.bozek@ifj.edu.pl}}
\affiliation{
Institute of Nuclear Physics, PL-31-342 Cracow, and
Institute of Physics, Rzeszow University, PL-35-959 Rzeszow, Poland}
\author{D. J. Dean\footnote{Electronic address~: deandj@ornl.gov}}
\affiliation{Physics Division, Oak Ridge National Laboratory, P.O. Box2008, Oak
Ridge, TN 37831-6373, USA}
\author{H. M\"uther\footnote{Electronic address~:
herbert.muether@uni-tuebingen.de}}
\affiliation{
Institut f\"ur Theoretische Physik, Universit\"at T\"ubingen, D-72076 T\"ubingen,
Germany}

\date{\today}

\begin{abstract}
We performed Self-Consistent Greens Function (SCGF) calculations 
for symmetric nuclear matter using realistic nucleon-nucleon (NN) 
interactions and effective low-momentum interactions 
($V_{low-k}$), which are derived from such realistic NN 
interactions. We compare the spectral distributions resulting from 
such calculations. We also introduce a density-dependent effective 
low-momentum interaction which accounts for the dispersive 
effects in the single-particle propagator in the medium. 
\end{abstract}


\maketitle

\section{Introduction}
The description of bulk properties of nuclear systems starting from realistic
nucleon-nucleon (NN) interactions is a long-standing and unsolved problem.
Various models for the NN interaction have been developed, which describe the
experimental NN phase shifts up to the threshold for pion production with high
accuracy\cite{cdbonn,arv18,nijmw,n3lo}. A 
general feature of all these interaction
models are strong short-range and tensor components, which lead to corresponding
correlations in the nuclear many-body wave-function. Hartree-Fock mean-field
theory, which represents the the lowest-order many-body calculations one
can perform with such realistic NN interactions, fails to produce bound
nuclei \cite{reviewartur,localint} precisely because Hartree-Fock does 
not fully incorporate many-body correlation effects.  

That correlations beyond the mean field are important is 
supported by experiments
exploring the spectral distribution of the single-particle strength. One 
experimental fact found in all nuclei is the global depletion 
of the Fermi sea.  A recent experiment from NIKHEF puts
this depletion of the proton Fermi sea in ${}^{208}$Pb at a little less than
20\%~\cite{bat01} in accordance with earlier nuclear matter
calculations~\cite{vond1}. Another consequence of the presence of short-range
and tensor correlations is the appearance of high-momentum components in the
ground state wave-function to compensate for 
the depleted strength of the mean field. Recent
JLab experiments~\cite{rohe:04} indicate that the amount and location of this
strength is consistent with earlier predictions for finite nuclei~\cite{mudi:94}
and calculations of infinite matter~\cite{frmu:03}.   

These data and their analysis, however, are not sufficient to allow for a
detailed comparison with the predictions derived from the various interaction
models at high momenta. In this paper, we want to investigate a possibility to 
separate the predictions for correlations at low and medium momenta, which are
constrained by the NN scattering matrix below pion threshold, from the high
momentum components, which may strongly depend on the underlying model for the
NN interaction. For that purpose we will perform nuclear many-body
calculations within a model space that allows for the explicit evaluation of
low-momentum correlations. The effective Hamiltonian for this model space will 
be constructed from a realistic interaction to account for for correlations
outside the model space. 

This concept of a model space and effective operators appropriately
renormalized for this model space has a long history in 
approaches to the nuclear many-body
physics. As an example we mention the effort to evaluate effective operators to
be used in Hamiltonian diagonalization calculations of 
finite nuclei. For a review on this topic
see e.g.~\cite{morten:04}. The concept of a model space for 
the study of infinite
nuclear matter was
used e.g. by Kuo et al.\cite{kumod1,kumod2,kumod3}. 
Also
the Brueckner-Hartree-Fock (BHF) approximation can be considered as a model
space approach. In this case one restricts the model space to just one
Slater-determinant and determines the effective interaction through 
a calculation of the G-matrix, the solution of the Bethe-Goldstone equation.  

The effective hamiltonians for such model space calculations have frequently
been evaluated within the Rayleigh-Schr\"odinger perturbation theory, leading to
a non-hermitian and energy-dependent result. The energy-dependence can be
removed by considering the so-called folded-diagrams as has been discussed e.g.
by Brandow\cite{brandow:67} and Kuo\cite{kuo:71}. We note that the
folded-diagram expansion yields effective interaction terms between three and
more particle, even if one considers a realistic interaction with two-body terms
only\cite{polls:83,polls:85}.

During the last years the folded-diagram technique has been applied 
to derive an effective low-momentum 
potential $V_{low-k}$\cite{bogner:03} from a realistic NN
interaction. By construction, $V_{low-k}$ potentials 
reproduce the deuteron binding-energy, the low-energy phase
shifts and the half-on-shell $T$ matrix calculated from the underlying realistic
NN interaction up to the chosen cut-off parameter.  The resulting $V_{low-k}$
turns out to be rather independent on the original NN interaction if this
cut-off parameter for the  relative momenta is below the value of the
pion-production threshold in NN scattering. 
The off-shell characteristics of the $V_{low-k}$ effective
interaction are not constrained by experimental data and can influence
the many-body character of the interaction. 

For finite nuclei we find 
that one does indeed obtain different binding energies 
for $^{16}$O depending on the 
underlying NN interaction from which one derives the $V_{low-k}$ 
interaction. For example, using coupled-cluster techniques 
at the singles and doubles level (CCSD) \cite{dean04} we find 
binding energies for $^{16}$O at a lab-momentum cutoff 
of $\Lambda=2.0$~fm$^{-1}$ to 
be $-143.4\pm 0.4$~MeV and $-153.3\pm 0.4$~MeV for the 
N$^3$LO \cite{n3lo} and CD-Bonn two-body interactions, respectively. 
The CCSD calculations were carried out at up to 7 major oscillator shells
(with extrapolations to an infinite model space) using the 
intrinsic Hamiltonian defined as $H=T-T_{cm}+V_{low-k}$ where 
$T_{cm}$ is the center of mass kinetic energy. 

Attractive energies are obtained if such a $V_{low-k}$ interaction is used in  a
Hartree-Fock calculation of nuclear matter or finite
nuclei\cite{corag:03,kuck:03}. High-momentum correlations, which are required to
obtain bound nuclear systems from a realistic NN interaction (see above) are
taken into account in the renormalization procedure which leads to $V_{low-k}$.
Supplementing these Hartree-Fock calculations with corrections up to third order
in the Goldstone perturbation theory leads to results for the ground-state
properties of $^{16}O$ and $^{40}Ca$, which are in fair agreement with the
empirical data\cite{corag:03}. (One should note that $T_{cm}$ was 
not included in these calculations.)
Calculations in infinite matter demonstrate that
$V_{low-k}$ seems to be quite a good approximation for the evaluation of
low-energy spectroscopic data. The results for the pairing derived from the bare
interaction are reproduced\cite{kuck:03}. The prediction of pairing properties
also agree with results obtained phenomenological interactions like the Gogny
force\cite{gogny,sedrak:03}. The $V_{low-k}$ interaction also yields a good
approximation for the calculated binding energy of nuclear matter at low
densities. 

At high densities, however, BHF calculations using $V_{low-k}$ yield
too much binding energy and do not reproduce the 
saturation feature of nuclear
matter\cite{kuck:03}. This is due to the fact that $V_{low-k}$ does not account
for the effects of the dispersive quenching of the two-particle propagator, as
it is done e.g. in the Brueckner $G$-matrix derived from a realistic NN
interaction. The saturation can be obtained if a three-body nucleon is added to
the hamiltonian\cite{bogner:05}.     

An alternative technique to determine an effective hamiltonian 
for a model space calculation is based on a unitary transformation of the
hamiltonian. It has been developed by Suzuki\cite{suzuki:82} and leads to an 
energy-independent, hermitian effective interaction.
The unitary-model-operator approach (UMOA) has also been used to evaluate the
ground-state properties of finite nuclei\cite{suz13,suz15,fuji:04,roth:05}.

In the present study we are going to employ the unitary transformation technique
to determine an effective interaction, which corresponds to the $V_{low-k}$
discussed above. This effective interaction will then be used in self-consistent
Green's function (SCGF) calculation of infinite nuclear matter. Various groups
have recently developed techniques to solve the corresponding equations and
determine the energy- and momentum-distribution of the single-particle strength
in a consistent way\cite{frmu:03,bozek0,bozek1,bozek2,dewulf:03,rd,frmu:05}.
Therefore we can study the correlation effects originating from $V_{low-k}$
inside the model space and compare it to the correlations derived 
from the bare interaction. 
Furthermore we use the unitary transformation technique 
to determine an effective
interaction which accounts for dispersive effects missing in the
original $V_{low-k}$ (see discussion above).

After this introduction we will present the method for evaluating the effective
interaction in section 2 and briefly review the basic features of the SCGF
approach in section 3. The results of our investigations are presented in
section 4, which is followed up by the conclusions.    

\section{Effective interaction}

For the definition and evaluation of an effective interaction to be used in a
nuclear structure calculation, which is restricted to a subspace of the Hilbert
space, the so-called model space, we follow the 
usual notation and define a projection operator $P$, which projects onto this
model space. The operator projecting on the complement of this subspace is
identified by $Q$ and these operators satisfy the usual relations like $P+Q=1$,
$P^2=P$, $Q^2=Q$, and $PQ=0=QP$. It is the aim of the Unitary Model Operator
Approach (UMOA) to define a unitary transformation $U$ in such a way, that the
transformed Hamiltonian does not couple the $P$ and $Q$ space, i.e.
$QU^{-1}HUP=0$. 

For a many-body system the resulting Hamiltonian can be evaluated in a cluster
expansion, which leads to many-body terms. This is very similar to the folded
diagram expansion, which has been discussed above. In UMOA studies of finite 
nuclei terms up to three-body clusters have been evaluated\cite{suz13,suz15}
indicating a convergence of the expansion up to this order.

In the present study we would like to determine an effective two-body
interaction and therefore consider two-body systems only. We 
define the effective
interaction as
\begin{equation}
V_{eff} = U^{-1}\left( h_0 + v_{12}\right) U - h_0\,,\label{eq:veff1}
\end{equation}
with $v_{12}$ representing the bare NN interaction. The operator $h_0$ denotes
the one-body part of the two-body system and contains the kinetic energy of the
interacting particles. This formulation 
will lead to an effective interaction
corresponding to $V_{low-k}$. Since, however, we want to determine an effective
interaction of two nucleons in the medium of nuclear matter, we will also
consider the possibility to add a single-particle potential to $h_0$. Note that
in any case $h_0$ commutes with the projection operators $P$ and $Q$.

The operator for the unitary transformation $U$ can be expressed as\cite{suz24}
\begin{equation}
U=(1+\omega-\omega ^{\dagger})(1+\omega \omega ^{\dagger} 
+\omega ^{\dagger}\omega )^{-1/2}\,,\label{eq:veff2}
\end{equation}  
with an operator $\omega$ satisfying $\omega=Q\omega P$ such that 
$\omega^2 = \omega^{\dagger 2} = 0$. In the following we will describe how to
determine the matrix elements of this operator $\omega$. As a first step we
solve the two-body eigenvalue equation
\begin{equation} 
\left( h_0 + v_{12}\right)\vert \Phi _{k}\rangle =E_{k}\vert \Phi _{k}\rangle\,.
\label{eq:veff2a}
\end{equation}
This can be done separately for each partial wave of the two-nucleon problem.
Partial waves are identified by total angular momentum 
$J$, spin $S$ and isospin $T$. The
relative momenta are appropriately discretized such that we can reduce the
eigenvalue problem to a matrix diagonalization problem. Momenta below the
cut-off momentum $\Lambda$ define the $P$ space and will subsequently be denoted
by $\vert p\rangle$ and $\vert p'\rangle$. Momenta representing the $Q$ space 
will be labeled by $\vert q\rangle$ and $\vert q'\rangle$, while states $\vert
i\rangle$, $\vert j\rangle$, $\vert k \rangle$ and  $\vert l \rangle$ refer to 
basis states of the total $P+Q$ space.

From the eigenstates $\vert \Phi _{k}\rangle$ we determine those $N_P$
($N_P$ denoting the dimension of the $P$ space) eigenstates 
$\vert \Phi _{p}\rangle$, which have the largest overlap with the $P$ space
and determine
\begin{equation}
\label{eq:veff3}
\langle q\vert\omega\vert p'\rangle =\sum_{p=1}^{N_P}\langle q\vert Q\vert
\Phi _{p}\rangle
\langle \tilde{\varphi}_{p}\vert p'\rangle,
\end{equation}    
with $\vert \varphi_{p}\rangle = P\vert \Phi_{p}\rangle$ and $\langle 
\tilde{\varphi}_{p}\vert$ denoting the biorthogonal state, satisfying
\begin{equation}
\sum_{p}\langle \tilde{\varphi} _{k}|p\rangle \langle p|\varphi _{k'}\rangle \quad
\mbox{and} \quad
\sum_{k}\langle p'|\tilde{\varphi} _{k}\rangle \langle \varphi _{k}|p\rangle
=\delta _{p,p'}\,.\label{eq:veff4}
\end{equation}
In the next step we solve the eigenvalue problem in the $P$ space
\begin{equation}
\omega ^{\dagger}\omega\vert\chi_{p}\rangle
=\mu _{p}^{2}|\chi_{p} \rangle\, ,\label{eq:veff5}
\end{equation}
and use the results to define 
\begin{equation}
\vert\nu _{p}\rangle =\frac{1}{\mu _{p}}\omega \vert\chi _{p}\rangle
,\label{eq:veff5a}
\end{equation}
which due to the fact that $\omega=Q\omega P$, can be written as
\begin{equation}
\label{eq:veff5b}
\langle q|\nu _{p}\rangle =\frac{1}{\mu _{p}}
\sum_{p'}\langle q|\omega |p'\rangle \langle p'|\chi_{p}\rangle\, .
\end{equation}
Using Eqs. (\ref{eq:veff5}) - (\ref{eq:veff5b}) and the representation of $U$ in Eq.
(\ref{eq:veff2}), the matrix elements of the unitary transformation operator $U$
can be written
\begin{eqnarray}
\label{eq:Up'p}
\langle p''|U|p'\rangle
&=&\langle p''|(1+\omega^{\dagger}\omega )^{-1/2}|p'\rangle \nonumber \\
&=&\sum_{p=1}^{N_P}(1+\mu_{p}^{2})^{-1/2}
\langle p''|\chi_{p}\rangle \langle \chi_{p}|p'\rangle \,,
\end{eqnarray}
\begin{eqnarray}
\label{eq:Uqp}
\langle q|U|p'\rangle
&=&\langle q|\omega (1+\omega^{\dagger}\omega )^{-1/2}|p'\rangle \nonumber \\
&=&\sum_{p=1}^{N_P}(1+\mu_{p}^{2})^{-1/2}\mu _{p}
\langle q|\nu _{p}\rangle \langle \chi _{p}|p'\rangle\, ,
\end{eqnarray}
\begin{eqnarray}
\label{eq:Upq}
\langle p'|U|q\rangle
&=&-\langle p'|\omega ^{\dagger}(1+\omega \omega ^{\dagger})^{-1/2}
|q\rangle \nonumber \\
&=&-\sum_{p=1}^{N_P}(1+\mu_{p}^{2})^{-1/2}\mu_{p}
\langle p'\vert\chi_{p}\rangle \langle \nu_{p}\vert q\rangle \,,  
\end{eqnarray}
\begin{eqnarray}
\label{eq:Uq'q}
\langle q'|U|q\rangle
&=&\langle q'|(1+\omega \omega ^{\dagger})^{-1/2}\vert q\rangle \nonumber \\
&=&\sum_{p=1}^{N_P}\{(1+\mu_{p}^{2})^{-1/2}-1\}
\langle q'|\nu _{p}\rangle \langle \nu _{p}|q\rangle + \delta
_{q,q'}\,.
\end{eqnarray}
These matrix elements of $U$ can then be used to determine the matrix elements
of the effective interaction $V_{eff}$ according to Eq.(\ref{eq:veff1}). They
might also be used to define matrix elements of other effective operators.

\section{Self-consistent Green's function approach}

One of the key quantities within the Self-consistent Green's Function (SCGF)
approach is the retarded single-particle (sp) Green's function or sp propagator
$G(k,\omega)$ (see e.g.\cite{diva:05}). Its imaginary part can be used to
determine the spectral function
\begin{equation}
\label{spec_g2}
A(k,\omega)=-2\,{\mathrm{Im}}\,G(k,\omega+{\mathrm{i}}\eta)\,.
\end{equation}
The spectral function provides the information about the energy- and
momentum-distribution of the single-particle strength, i.e. the probability for
adding or removing a particle with momentum $k$ and leaving the residual system
at an excitation energy related to $\omega$.   
In the limit of the mean-field or quasi-particle approximation the spectral
function is represented by a $\delta$-function and takes the simple form
\begin{equation}
A(k,\omega)=2\pi\delta(\omega -\varepsilon_k) 
\,,\label{eq:specqp}
\end{equation}
with the quasi-particle energy $\varepsilon_k$ for a particle with momentum $k$.
The sp Green's function can be obtained from the solution of the Dyson equation,
which reduces for the system of homogeneous infinite matter to a a simple
algebraic equation
\begin{equation}
\left[\omega -\frac{k^2}{2m}-\Sigma(k,\omega)\right] 
G(k,\omega) = 1\,,\label{eq:dyson}
\end{equation}
where $\Sigma(k,\omega)$ denotes the complex self-energy. The self-energy can be
decomposed into a generalized Hartree-Fock part plus a dispersive contribution
\begin{equation}
\label{spec_Sigma}
\Sigma(k,\omega)=\Sigma^{HF}(k)-\frac{1}{\pi}\int_{-\infty}^{+\infty}
{\mathrm{d}}\omega^{\prime} \, \frac{{\mathrm{Im}}\Sigma(k,\omega^{\prime}+
{\mathrm{i}}\eta)}
{\omega-\omega^{\prime}}.
\end{equation}
The next step is to obtain the self energy in terms of the in-medium
two-body scattering $T$ matrix. It is possible to express
${\mathrm{Im}}\Sigma(k,\omega+{\mathrm{i}}\eta)$ in terms of the
retarded $T$ matrix~\cite{frmu:03,bozek3,kadanoff}
 (for clarity, spin- and isospin quantum number
are suppressed)
\begin{eqnarray}
\label{im_sigma}
{\mathrm{Im}}\Sigma(k,\omega+{\mathrm{i}}\eta)&=&
\frac{1}{2}\int \frac{{\mathrm{d}}^3k^{\prime}}{(2\pi)^3}
\int_{-\infty}^{+\infty} \frac{{\mathrm{d}}\omega^{\prime}}{2\pi}
\left<{\mathbf{kk}}^{\prime}|
{\mathrm{Im}}T(\omega+\omega^{\prime}+{\mathrm{i}}\eta)|
{\mathbf{kk}}^{\prime}\right> \nonumber \\ && \qquad \times
[f(\omega^{\prime})+b(\omega+\omega^{\prime})]
A(k^{\prime},\omega^{\prime}).
\end{eqnarray}
Here and in the following $f(\omega)$ and $b(\omega)$ denote the Fermi and Bose
distribution functions, respectively. These 
functions depend on the chemical potential
$\mu$ and the inverse temperature $\beta$ of the system. The in-medium
scattering matrix $T$ is to be determined as a 
solution of the integral equation
\begin{eqnarray}
\left<{\mathbf{kk}}^{\prime}|T(\Omega+{\mathrm{i}}\eta)|
{\mathbf{pp}}^{\prime}\right> & = &\left<{\mathbf{kk}}^{\prime}|V|
{\mathbf{pp}}^{\prime}\right> + \int
\frac{d^3q\,d^3q^\prime}{\left(2\pi\right)^6} \left<{\mathbf{kk}}^{\prime}|V|
{\mathbf{qq}}^{\prime}\right>G^0_{\mathrm{II}}(\mathbf{qq}^\prime,\Omega+i\eta)
\nonumber \\ &&\quad\quad\times
\left<{\mathbf{qq}}^{\prime}|T(\Omega+{\mathrm{i}}\eta)|
{\mathbf{pp}}^{\prime}\right>\,,\label{eq:tscat0}
\end{eqnarray}
where
\begin{equation}
\label{two_pp}
G^0_{\mathrm{II}}(k_1,k_2,\Omega+i\eta)=
\int_{-\infty}^{+\infty}\frac{{\mathrm{d}}\omega}{2\pi}
\int_{-\infty}^{+\infty}\frac{{\mathrm{d}}\omega^{\prime}}{2\pi}
A(k_1,\omega)A(k_2,\omega^{\prime})
\frac{1-f(\omega)-f(\omega^{\prime})}
{\Omega-\omega-\omega^{\prime}+i\eta}\,.
\end{equation}
stands for the 
two-particle Green's function of two non-interacting but
dressed nucleons. The matrix elements of the two-body interaction $V$ represent
either the bare NN interaction $v_{12}$ or the effective interaction $V_{eff}$,
in which case the integrals are cut at the cut-off parameter $\Lambda$.

The in-medium scattering equation (\ref{eq:tscat0}) can be reduced to a set of
one-dimensional integral equations if the two-particle Green's function in 
(\ref{two_pp}) is written as a function of the total and relative momenta of 
the interacting pair of nucleons and the usual angle-average approximation is
employed (see \textit{e.g.}~\cite{angleav} for the accuracy of this 
approximation). This leads to integral equations in the usual partial waves,
which can be solved very efficiently if the two-body interaction is represented
in terms of separable interaction terms of a sufficient rank\cite{bozek1}. 

Finally, we consider the generalized Hartree-Fock contribution to the
self-energy in (\ref{spec_Sigma}), which takes the form 
\begin{equation}
\label{hf_sigma}
\Sigma^{HF}(k)
=
\frac{1}{2} \int \frac{{\mathrm{d}}^3k^{\prime}}{(2\pi)^3}   
\left<{\mathbf{k}},{\mathbf{k}}^{\prime}\right|
V \left|{\mathbf{k}},{\mathbf{k}}^{\prime}\right> n(k^{\prime}),
\end{equation}
where $n(k)$ is the correlated momentum distribution, which is to be calculated
from the spectral function by
\begin{equation}
\label{occupation}
n(k)=
\int_{-\infty}^{+\infty} \frac{{\mathrm{d}}\omega}{2\pi}
f(\omega)
A(k,\omega).
\end{equation}
Also the energy per particle, $E/A$, can be calculated from the spectral
function using Koltun's sum
rule
\begin{equation}
\label{eda}
\frac{E}{A}=\frac{1}{\rho}
\int \frac{{\mathrm{d}}^3k}{(2\pi)^3}
\int_{-\infty}^{+\infty} \frac{{\mathrm{d}}\omega}{2\pi}
\frac{1}{2}\left(\frac{k^2}{2m}+\omega\right)A(k,\omega)f(\omega)\,.
\end{equation}
 Eqs.(\ref{spec_g2})-(\ref{occupation}) define the so-called $T$-matrix approach
to the SCGF equations. They form a symmetry conserving approach in the sense of
\cite{kadanoff}, which means that thermodynamical relations like the
Hughenholtz-Van Hove theorem\cite{hugenholtz,bozek1} are obeyed.

The Brueckner-Hartree-Fock (BHF) approximation, which is very popular in nuclear
physics, can be regarded as a simple approximation to this $T$-matrix approach.
In the BHF approximation one reduces the spectral function $A(k,\omega)$ to the
quasiparticle approximation (\ref{eq:specqp}). Furthermore one ignores the
hole-hole scattering terms in the scattering Eq.(\ref{eq:tscat0}), which means
that one replaces
\begin{equation}
\left(1-f(\omega)-f(\omega')\right) \quad\rightarrow \quad \left(1-f(\omega)
\right) \left(1-f(\omega')\right)\,,
\label{eq:pauliop}
\end{equation} 
which is the usual Pauli operator (at finite temperature). This reduces the
in-medium scattering equation to the Bethe-Goldstone equation. The removal of
the hole-hole scattering terms leads 
to real self-energies $\Sigma(k,\omega)$ at
energies $\omega$ below the chemical potential, i.e.~for the hole states. 

\section{Results and discussion}

In the following we discuss results for symmetric nuclear matter
obtained from Self-Consistent Greens Function (SCGF) calculations. These
calculations are either performed in the complete Hilbert space using the bare
CD-Bonn \cite{cdbonn} interaction or in the model space, which is defined by a
cut-off parameter $\Lambda$ = 2 fm$^{-1}$ in the two-body scattering equation,
employing the corresponding effective interaction $V_{low-k}$, which is derived
from the CD-Bonn interaction using the techniques described in Sect II. 
We note that
using this unitary model operator technique we were able 
to reproduce the results of the BHF calculations presented in \cite{kuck:03},
which used tabulated matrix elements of \cite{bogner:03}, with good accuracy.
The NN interaction has been restricted to partial waves with total angular
momentum $J$ less than 6.

\begin{figure}      
\begin{center}
\includegraphics[width=10.5cm]{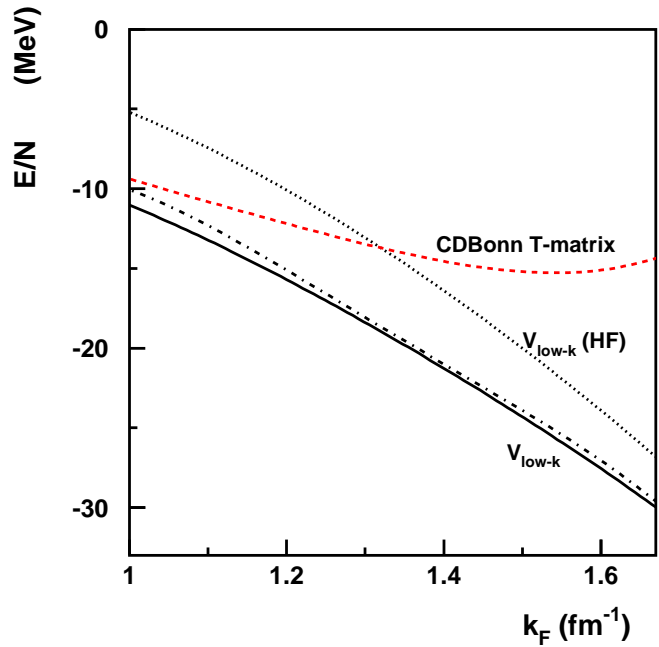}
\end{center}
\caption{(Color online) Binding energy per nucleon for symmetric nuclear matter
as function of the Fermi momentum: Results of self-consistent $T$-matrix
calculations for the CD-Bonn potential (dashed line), are compared to results of
calculations using $V_{low-k}$ with  $\Lambda=2$fm$^{-1}$ in the Hartree-Fock
approximation (dotted line), the self-consistent second order approximation
(dashed-dotted line) and for the self-consistent $T-$matrix approximation (solid
line) within the model space. }
\label{fig:becd1}
\end{figure}

Results for the calculated energy per nucleon are displayed in
Fig.~\ref{fig:becd1} for various densities, which are labeled by the
corresponding Fermi momentum $k_F$. The effective interaction $V_{low-k}$
accounts for a considerable fraction of the short-range NN correlations, which
are induced by realistic interactions like the CD-Bonn interactions. Therefore,
already the Hartree-Fock approximation using this $V_{low-k}$ yields 
reasonable results for the energies as can be seen from the dotted line of 
Fig.~\ref{fig:becd1}. Hartree-Fock calculations using the bare CD-Bonn
interaction yield positive energies ranging between 2 MeV per nucleon and 15 MeV
per nucleon for the densities considered in this figure. Note that the CD-Bonn
interaction should be considered as a soft realistic interaction. Interaction
models, which are based on local potentials, like the Argonne interaction 
\cite{arv18}, yield more repulsive Hartree-Fock energies \cite{localint}.

The inclusion of correlations within the model space yields a substantial
decrease of the energy. The self-consistent $T$-matrix approach provides
additional attraction ranging between 6 MeV per nucleon at a density of 
0.4 $\rho_0$ (with $\rho_0$ the empirical saturation density) and 3 MeV per
nucleon at 2 $\rho_0$. The fixed cut-off parameter $\Lambda$ seems to reduce
the phase-space available for correlations beyond the mean-field approach 
at higher densities. Therefore the energy calculated in the self-consistent 
$T$-matrix approach reduces to the Hartree-Fock result at large densities.

Fig.~\ref{fig:becd1} also displays the energies resulting from a SCGF
calculation within the model space, in which the $T$-matrix has been
approximated by the corresponding scattering matrix including only terms up to
second order in the NN interaction $V$. The results of such second-order
calculations in $V_{low-k}$ are represented by the dashed-dotted line and show a
very good agreement with the model-space calculations including the full 
$T$-matrix. This confirms the validity of approaches, which consider correlation
effects within the model-space in a perturbative way.

All these model space calculations using $V_{low-k}$, however, fail to 
reproduce the results of the SCGF calculations, which are obtained in the 
complete space using the bare NN interaction, which are labeled by CD Bonn
T-matrix in Fig.~\ref{fig:becd1}. In particular, the model space calculations
yield to attractive energies at high densities and therefore do not exhibit a
minimum for the energy as a function of density.
This confirms the results of the BHF calculations of \cite{kuck:03}.     

It has been argued \cite{kuck:03} that this 
overestimate of the binding energy at high densities
is due to the fact that $V_{low-k}$ does not account for the quenching of
correlation effects, which is due to the Pauli principle and the dispersive
effects in the single-particle propagator getting more 
important with increasing
density. Therefore we try to account for the dispersive quenching effects by
adopting the following two-step procedure. 

In a vein similar to the use of a G-matrix within a self-consistent BHF 
calculation, as a 
first step we perform BHF calculations using $V_{low-k}$. The resulting
single-particle spectrum is approximated by an effective 
mass parameterization.
This parameterization of the mean field is employed to 
define the single-particle
operator $h_0$, used in Eq.~(\ref{eq:veff1}) and the following equations of
Sect.~II (see also \cite{fuji:04}). The resulting effective interaction is used
again for a BHF calculation within the model space, leading to an update of 
the mean field parameterization. The procedure is repeated until a
self-consistent result is obtained. Since the mean field parameterization
depends on the density, this method yields an effective density-dependent
interaction, which in the limit of the density $\rho\to 0$ coincides with 
$V_{low-k}$. Therefore we call this effective interaction the density dependent
$V_{low-k}$ or in short $V_{low-k}(\rho)$. Such a procedure amounts 
to summing up certain higher order terms in the full many-body problem. 

\begin{figure}      
\begin{center}
\includegraphics[width=10.5cm]{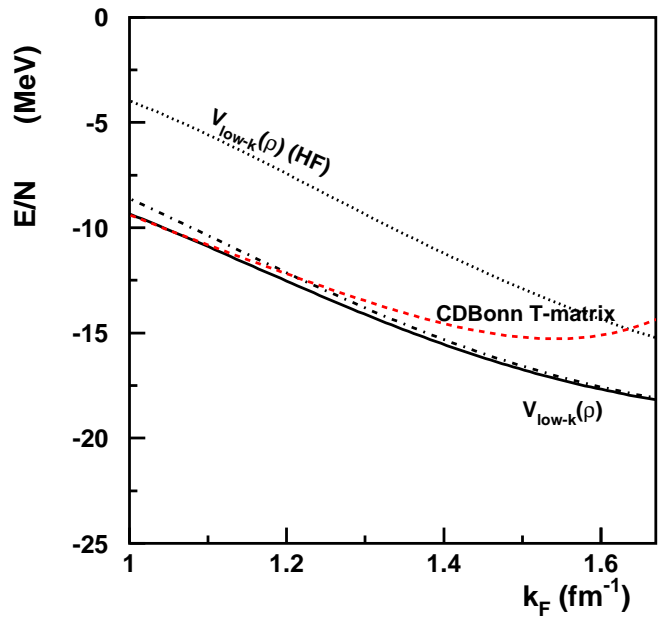}
\end{center}
\caption{(Color online)
Same as Fig.~\ref{fig:becd1} but for $V_{low-k}(\rho)$ calculated at each density}
\label{fig:becd2}
\end{figure}

In a second step this $V_{low-k}(\rho)$ is used in SCGF calculations at the
corresponding density. Energies resulting from such model space calculations
using $V_{low-k}(\rho)$ are presented in Fig.~\ref{fig:becd2}. The comparison of
the various calculations within the model space exhibits the same features as
discussed above for the original $V_{low-k}$. The correlation within the model
space provide a substantial reduction of the energy as can be seen from the
comparison of the self-consistent $T$-matrix approach with the Hartree-Fock
results. The approach treating correlations up to second order in 
$V_{low-k}(\rho)$ yields energies which are very close to the complete $T$-matrix
approach.

The density dependence of the effective interaction $V_{low-k}(\rho)$ yields a
significant improvement for the comparison between the model space calculations
and the SCGF calculation using the bare CD-Bonn interaction. Note that the
energy scale has been adjusted going from Fig.~\ref{fig:becd1} to 
Fig.~\ref{fig:becd2}. The discrepancy remaining at densities above $\rho_0$ might
be due to the effects of the Pauli quenching, which are not included in 
$V_{low-k}(\rho)$. These deviations could also originate from the simple
parameterization of the dispersive quenching in $V_{low-k}(\rho)$. 

\begin{figure}      
\begin{center}
\includegraphics[width=10.5cm]{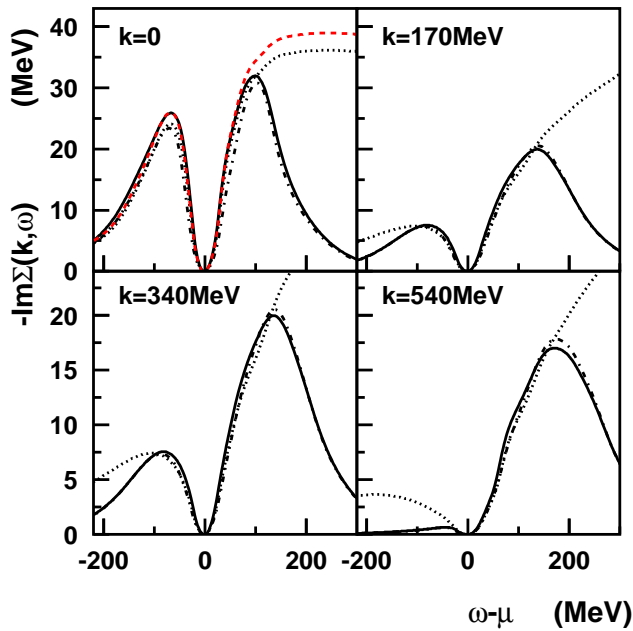}
\end{center}
\caption{(Color online)  Imaginary part of the self-energy as a function of the energy
$\omega$ for various momenta $p$ as indicated in the panels (see Eq.~(
\protect{\ref{im_sigma}})). The results have been determined for the empirical
saturation density
 $\rho_0$; using $V_{low-k}$ in the $T$-matrix approximation (solid line), using
$V_{low-k}$ in the second order approximation (dashed-dotted line), and employing
CD-Bonn interaction in the $T$-matrix approximation (dotted line). The dashed
line in the first panel 
denotes the results of the $T$-matrix calculation with the CD-Bonn potential
rescaled by the ratio of the effective masses at the Fermi momentum 
obtained with the $V_{low-k}$ and the bare CD-Bonn potential. 
}
\label{fig:im10}
\end{figure}

Our investigations also provide the possibility to explore the effects of correlations
evaluated within the model space using the effective interaction $V_{low-k}$. We can
furthermore compare these correlation effects with the corresponding effects determined
by the bare interaction in the unrestricted space. As a first example, we discuss the
imaginary part of the self-energy calculates at the empirical saturation density
$\rho_0$ for various nucleon momenta $p$ as displayed in Fig.~\ref{fig:im10}. 
The calculations within the model space
reproduce the results of the unrestricted calculations with a good accuracy in the
energy interval for $\omega$ ranging between 50 MeV below and 50 MeV above the
chemical potential $\mu$. The remaining differences around the Fermi energy
can be attributed to the difference in the effective masses obtained 
using the $V_{low-k}$ 
and the bare potential \cite{wi98}. The agreement between the
$T$-matrix results around $\omega=\mu$ using the two potentials is 
improved if one rescales by the
ratio of the effective masses.
The imaginary part calculated with $V_{low-k}$, however, 
is much smaller than the corresponding result obtained 
for the bare interaction at
energies $\omega -\mu$ above 100 MeV. Furthermore the model 
space calculation do not
reproduce the imaginary part for energies below 
the chemical potential at momenta $k$
above 400 MeV/c. 

\begin{figure}      
\begin{center}
\includegraphics[width=10.5cm]{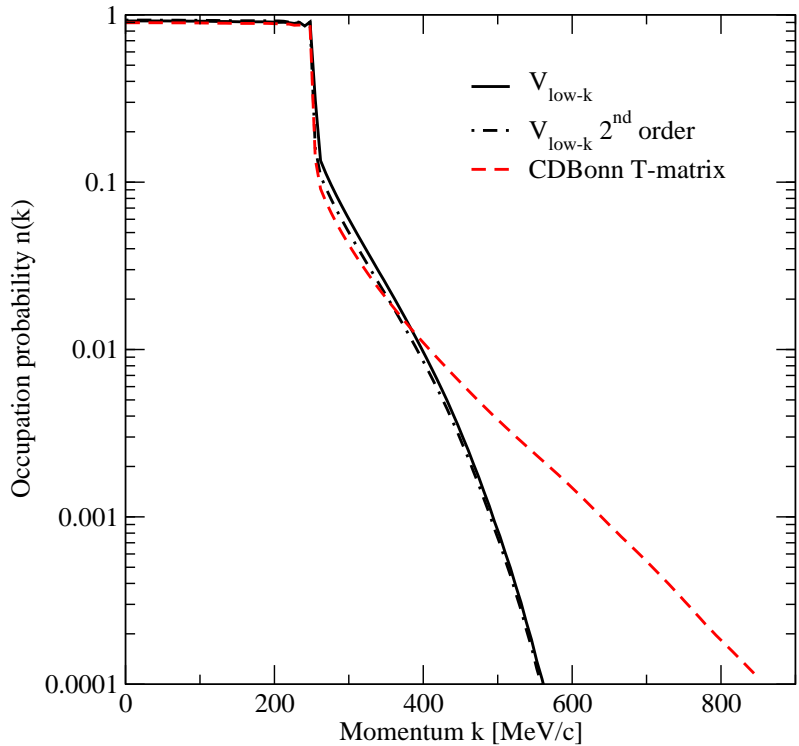}
\end{center}
\caption{(Color online) Momentum distribution $n(k)$ (see
Eq.~(\protect{\ref{occupation}})) calculated for nuclear matter at the empirical
saturation density $\rho_0$. Results of the $T$-matrix approximation within the model
space (solid line) are compared to results of the second order approximation
(dashed-dotted line) and the $T$-matrix approximation (dotted line) in the
unrestricted space.}
\label{fig:nofk}
\end{figure}

The imaginary part of the self-energy is a very important ingredient for the
evaluation of the spectral function $A(k,\omega )$ and therefore also for the
calculation of the occupation probability $n(k)$ (see Eq.~(\ref{occupation})). The
small values for the imaginary part of the self-energy at high momenta $k$ and
negative energies $\omega -\mu$ leads to occupation probabilities at these momenta,
which are much smaller than the corresponding predictions derived from bare realistic
NN interactions as can be seen from Fig.~\ref{fig:nofk}. This missing strength in the
prediction of $V_{low-k}$ at high momenta is accompanied by larger occupation
probabilities at low momenta. The self-consistent $T$-matrix approximation
using CD-Bonn yields an occupation probability at $k=0$ of 0.897, while the
corresponding number using $V_{low-k}$ is 0.920. At this density, the calculation
including only terms up to second order in $V_{low-k}$ yields a rather good
approximation to the self-consistent $T$-matrix approximation within the model space.

\begin{figure}      
\begin{center}
\includegraphics[width=10.5cm]{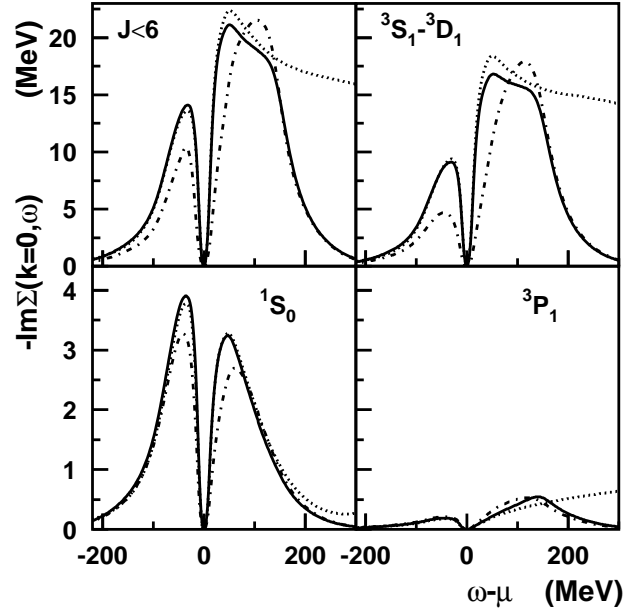}
\end{center}
\caption{Imaginary part of the self-energy as a function of the energy
$\omega$ for nucleons with momentum $k=0$ calculated at the density $\rho=0.4\times
\rho_0$. Results of the $T$-matrix approach (solid line) and the second order
approximation (dashed-dotted line) within the model space are compared to results
obtained in the unrestricted calculation (dotted line).}
\label{fig:im04}
\end{figure}

As a second example we consider the imaginary part of the self-energy calculated at a
lower density $\rho=0.4\times \rho_0$. The results displayed in  Fig.~\ref{fig:im04}
refer to nucleons with momentum $k=0$. Also at this density we find that the imaginary
part evaluated with $V_{low-k}$ drops to zero at large positive energies much faster
than the predictions derived from the bare interaction (see upper panel on the left in
Fig.~\ref{fig:im04}). 

It is worth noting, that at this low density the second order approximation is not 
such a good approximation to the full $T$-matrix approach as it is for the higher
densities. Characteristic differences between the dashed-dotted and the solid line show
up at energies $\omega$ close to the chemical potential. In order to trace the origin 
of these differences we display in Fig.~\ref{fig:im04} the contributions of various
partial waves of NN interaction channels to this imaginary part. It turns out that the
differences are largest in the $^3S_1-^3D_1$ and the $^1S_0$ channels. This means that
the perturbative approach is not very successful in those two channels which tend to
form quasi-bound states. In these channels all particle-particle hole-hole ladders
have to be summed up to obtain the pairing solution. Note, that the pairing solutions
are suppressed at higher densities, if the effects of short-range correlations are
properly taken into account\cite{bozek4,muwi05}.  

Furthermore we would like to point out that a different scale is used in the two
lower panels of Fig.~\ref{fig:im04}. Taking this into account it is evident from this
figure that the main contribution to the imaginary part of the self-energy, and that
means the main contribution to the character of the deviation of the spectral function
from the mean-field approach originates from the NN interaction in the $^3S_1-^3D_1$
channel.  

\section{Conclusions}

During the last few years it has become very 
popular to perform nuclear structure
calculations using effective low-momentum NN interactions. These $V_{low-k}$
interactions are based on a realistic model of the NN interaction. They are
constructed to be different from zero only within a model space 
defined by a cut-off $\Lambda$ in the relative momenta of 
the interacting 
nucleons. Within this model space they reproduce the NN 
data of the underlying bare interaction, although the 
many-body solutions may show differences with different 
starting NN interactions. 

For this study we performed Self-Consistent Greens Functions (SCGF) 
calculations of symmetric nuclear matter employing $V_{low-k}$
effective interactions as well as the bare
CD Bonn interaction they are based on. Special attention 
was paid to the correlations
which can be described within this model space 
as compared to correlations predicted
by the underlying interaction within the unrestricted space. 

Using a cut-off $\Lambda$ = 2 fm$^{-1}$ we find that the 
spectral distribution of 
the single-particle strength in an energy window 
of plus minus 50 MeV around the Fermi
energy is rather well reproduced by the calculation 
using $V_{low-k}$. The effective
interaction $V_{low-k}$ is softer than typical realistic NN 
interactions. Therefore
for many observables it is sufficient to approximate the 
full in-medium scattering
matrix $T$ by the approximation including terms up to second order in
$V_{low-k}$. This justifies the use of the resummed effective interaction in
many-body approximations that do not the include ladder-diagram resummation.
Special attention must be paid to nuclear systems at smaller 
densities: the possible
formation of quasi-bound states may require the 
non-perturbative treatment of the NN
scattering in the medium. This also has implications for the use
of $V_{low-k}$ in studies of weakly bound nuclear systems. 

The model space approach cannot reproduce correlation effects, 
which lead to spectral
strength at high energies and high momenta. For nuclear matter 
at the empirical 
saturation density $\rho_0$ the momentum distribution is reliably 
predicted up to a
momentum of 400 MeV/c.

The $V_{low-k}$ approach overestimates the 
binding energy per nucleon at high densities. Therefore we introduced a
density-dependent effective interaction $V_{low-k}(\rho)$ 
which we constructed along the same line as the original $V_{low-k}$. The
new effective interaction
accounts for a dispersive correction of the single-particle propagator 
in the medium.
This improves the behavior of the effective interaction 
significantly. For densities
above $\rho_0$, however, the binding energies calculated 
with $V_{low-k}(\rho)$ are 
still too large. This might be improved by 
determining effective three-nucleon forces
explicitly from the underlying bare interaction.
       
\acknowledgments
This work is supported in part by the Polish State Committee for Scientific
Research Grant No. 2P03B05925U.S, the Department of Energy 
under Contract Number DE-AC05-00OR22725 with UT-Battelle, LLC (Oak 
Ridge National Laboratory) and the Deutsche Forschungsgemeinschaft (SFB 382).

\end{document}